\documentclass{article}

\usepackage{bm}
\usepackage{graphicx}
\usepackage{cite}
\usepackage{amsgen,amsfonts,amssymb,amsbsy}

\newcommand{\NAME}[1]{\textsc{#1}}

\newcommand{\BY}[1]{\NAME{#1},}
\newcommand{\IN}[4]{\textit{#1}, \textbf{#2} (#3) #4}

\newcommand{\TITLE}[1]{\textit{#1}}

\begin{document}
\title{Introduction to the theory of x-ray matter interaction}
\author{Nina Rohringer
        \thanks{nina.rohringer@desy.de}\\
        DESY, Deutsches Elektronen-Synchrotron, 22607 Hamburg\\
        Center for Free Electron Laser Science, 22607 Hamburg
        }

\maketitle
\begin{abstract}
This contribution summarises the theoretical basics for the theory of x-ray matter interaction and is based on a quantised description of the electromagnetic field. In view of a formally transparent presentation, we adopt the approach of second quantisation for the electronic degrees of freedom. We start out from the minimal coupling Hamiltonian that in principle contains all levels of electron correlation and treat the x-ray matter interaction in lowest order perturbation theory. We give the explicit derivation of the photoionisation (x-ray absorption) cross section and the doubly differential cross sections for elastic and inelastic Thomson scattering, as well as resonant inelastic x-ray scattering. A connection to relevant experimental techniques is attempted. At some instances, the electronic Hamiltonian is approximated by a mean-field Hamiltonian, in order to facilitate the interpretation of the results and connect experimental observables with the underlying fundamental quantities of the quantum system under study. 
\end{abstract}

\section{Introduction}
These lecture notes aim to present a summary of the theory of x-ray matter interaction in a concise and yet broad enough way to cover all the different phenomena of x-ray matter interaction. Following an approach that was summarised by Robin Santra in his PhD tutorial \cite{santra} on "Concepts in X-ray Physics", I introduce x-ray matter interaction from a very fundamental derivation based on the minimal coupling Hamiltonian of non-relativistic quantum electrodynamics, and in representing the many-body electronic Hamiltonian in terms of second quantisation. The details of the derivation are well summarised in R. Santra's tutorial and I invite the interested student to an in-depth study of that work. In that work, fundamental x-ray matter interaction processes such as photoionisation, elastic x-ray scattering and Compton scattering are discussed and the cross section of these processes are derived. The work also derives relaxation mechanisms following the creation of an inner-shell vacancy, such as the electronic Auger decay and the radiative decay (fluorescence). 
Here, I summarise that work in concise, self-consistent form and extend the presentation to a few additional processes that find application in present day measurement techniques. In particular, I am complementing the tutorial by including the process of resonant inelastic x-ray scattering (RIXS), and inelastic Thomson x-ray scattering.\\
This article is structured as follows: In chapter 2 we introduce the minimal coupling Hamiltonian that is the basis for the interaction of the matter degrees of freedom with the electromagnetic field. Chapter 3 is a summary of perturbation theory in the interaction picture and we define the transition rate up to second order interaction with the electromagnetic field. In chapter 4 we discuss x-ray absorption and shortly connect our theoretical findings to photoabsorption spectroscopy. Chapter 5 gives an introduction to coherent elastic scattering, establishing the basic equation that connects the electronic structure to the elastic scattering cross section. In chapter 6 we introduce inelastic Thomson scattering and make a connection to the dynamic structure factor and the time-dependent pair-distribution function of disordered materials. We also discuss the basics of resonant inelastic x-ray scattering and give a general expression for the doubly differential inelastic scattering cross section.\\
Atomic units are employed, i.e., $m_e=1$, $|e|=1$, $\hbar = 1$, and $c=1/\alpha$, where $m_e$ is the electron mass, 
$e$ is the electron charge, $\hbar$ is Planck's constant divided by $2\pi$, $c$ is the speed of light in vacuum, and 
$\alpha = \frac{e^2}{\hbar c} \approx 1/137$ is the fine-structure constant.  The atomic unit of length is the bohr, 
$a_0 = \frac{1}{\alpha}\frac{\hbar}{m_e c} \approx 0.529$~\AA.  The atomic unit of cross section is $a_0^2 \approx 28.0$~Mb, 
where a {\em barn} (b) equals $10^{-28}$~m$^2$. The atomic unit of energy is the hartree, $E_h = m_e c^2 \alpha^2 \approx 27.2$~eV.
\section{The Matter-Field Hamiltonian}
\subsection{The Hamiltonian of the electromagnetic field}
In the formulation of quantum electrodynamics the electromagnetic field is quantised. An extended theoretical treatment of the quantisation of the electromagnetic field can be found in various text books (see for example Chapter 2 of reference \cite{heitler}). Here we shall give a short summary of the results of quantisation. We consider a pure radiation field, that is composed by only transverse electric and magnetic fields. The radiation field is then characterised by a vector potential ${\bm A}({\bm x})$ that is expanded in modes of wave vector ${\bm k}$ and polarisation $\lambda$ in an imaginary box of volume $V$.  We adopt the Coulomb gauge ${\bm \nabla}\cdot{\bm A}=0$. For a radiative field (no static electric and magnetic fields) in Coulomb Gauge, the scalar potential is zero. In its quantised form, the vector potential reads
\begin{equation}
\label{vectorpotential}
\hat{{\bm A}}({\bm x}) = \sum_{{\bm k},\lambda} \sqrt{\frac{2\pi}{V \omega_{\bm k} \alpha^2}}
\left\{\hat{a}_{{\bm k},\lambda} {\bm \epsilon}_{{\bm k},\lambda} {\rm e}^{{\rm i}{\bm k}\cdot{\bm x}}
+ \hat{a}^{\dag}_{{\bm k},\lambda} {\bm \epsilon}^{\ast}_{{\bm k},\lambda} {\rm e}^{-{\rm i}{\bm k}\cdot{\bm x}}\right\}\;.
\end{equation}
Here $\hat{a}^\dag_{{\bm k},\lambda}$ ($\hat{a}_{{\bm k},\lambda}$) creates (annihilates) a photon in the mode $({\bm k},\lambda)$. The creation and annihilation operators of the photon field follow the common bosonic commutation relations:
\begin{equation}
[\hat{a}_{{\bm k},\lambda},\hat{a}_{{\bm k}',\lambda'}]=
[\hat{a}^{\dag}_{{\bm k},\lambda},\hat{a}^{\dag}_{{\bm k}',\lambda'}]= 0\;\;
\mathrm{and}\;\;
[\hat{a}_{{\bm k},\lambda},\hat{a}^{\dag}_{{\bm k}',\lambda'}] = \delta_{{\bm k},{\bm k}'}\delta_{\lambda,\lambda'}\;.
\end{equation}
In terms of this mode expansion the Hamiltonian of the electromagnetic field is given by
\begin{equation}
\hat E_{EM} = \sum_{ {\bm k},\lambda} \omega_{\bm k} \hat{a}^\dag_{{\bm k},\lambda} \hat{a}_{{\bm k},\lambda},
\end{equation}
where we omitted the vacuum contribution of the field.
\subsection{The principle of minimal coupling and the molecular Hamiltonian}
To describe the interaction of matter with the electromagnetic field we adopt the minimal coupling Hamiltonian.
In its most fundamental form, the minimal coupling Hamiltonian is given by a sum over single-particle Hamiltonians $\hat{h}_i$ of all the matter constituents $i$ of charge $q_i$ and mass $m_i$ (in our case nuclei and electrons) of the form
\begin{equation}
\label{eq1}
\hat{h}_i = \frac{[\hat{{\bm p}}_i -\alpha q_i \hat{{\bm A}}({\bm x}_i)]^2}{2 m_i} + q_i \Phi_i({\bm x}_i).
\end{equation}
Since we adopt a non-relativistic description of the electronic degrees of freedom, spin-dependent coupling terms, such as spin-orbit coupling, would need to be added to the Hamiltonian. For the current treatment, we however disregard those spin-related coupling terms.  Applying Gauss' law for a system of point charges, the scalar potential $\Phi({\bm x})$ is determined by
\begin{equation}
\label{eq8}
\Phi_i({\bm x})= \sum_{j\neq i} \frac{q_j}{|{\bm x} - {\bm x}_j|},
\end{equation}
where the sum runs over all charged particles and represents the total potential resulting from the Coulomb interaction between all point particles, including electron-electron and electron-nucleus interaction.\\
Without a radiation field, i.e., $\hat{\bm A}=0$, the Hamiltonian $\hat{H}_0:=\sum_i\hat{h_i}$ reduces to the
pure molecular Hamiltonian 
\begin{equation}
\hat{H}_{\mathrm{mol}} = \hat{T}_{\mathrm{N}} + \hat{V}_{\mathrm{NN}} + \hat{H}_{\mathrm{el}},
\end{equation}
given by the sum of the kinetic energy of the nuclei $\hat{T}_{\mathrm{N}}$, the potential energy of the nuclei  $\hat{V}_{\mathrm{NN}}$ due to their Coulomb repulsion and the remaining electronic Hamiltonian $\hat{H}_{\mathrm{el}}$. Explicitly $\hat{T}_{\mathrm{N}}$ is given by
\begin{equation}
\label{eq10}
\hat{T}_{\mathrm{N}} = -\frac{1}{2} \sum_n \frac{\nabla^2_n}{M_n}\;\mbox{ and}
\end{equation}
\begin{equation}
\label{eq11}
\hat{V}_{\mathrm{NN}} = \sum_{n < n'} \frac{Z_n Z_{n'}}{|{\bm R}_n - {\bm R}_{n'}|}
\end{equation}
is the potential energy due to nucleus-nucleus repulsion.  Here, ${\bm R}_n$, $M_n$, and $Z_n$ are, respectively, the position, mass, and charge of the $n$-th nucleus. The electronic Hamiltonian is given by
\begin{equation}
\label{hel}
\hat{H}_{\mathrm{el}} = -\frac{1}{2} \sum_{i} \nabla_i^2
- \sum_{n,i}\frac{Z_n}{|{\bm x_i} -{\bm R}_n|}\\
+ \frac{1}{2} \sum_{i\neq j} \frac{1}{|{\bm x_i} - {\bm x_j}|} \nonumber
\end{equation}
where the sums extend over all electrons $i,j$ and all nuclei $n$. Besides relativistic effects, and effects such as spin-orbit coupling, the Hamiltonian of the molecular system of eq.\ (\ref{hel}) is exact. We do not consider the interaction of the electromagnetic waves with the nuclei, since compared to the electrons, they are too heavy to show substantial direct coupling to the field.\\
The interaction Hamiltonian between electrons and the field is
\begin{eqnarray}\label{hint0}
\hat{H}_{{\rm int}}=\sum_i^{N_{\rm el}} \left[ \alpha \hat{\bm A}(\bm x_i)\cdot\frac{\bm \nabla_i}{{\rm i}}+\frac{\alpha^2}{2}\hat{\bm A}^2(\bm x_i)\right]\;\;,
\end{eqnarray}
where the sum runs over all electrons $N_{\rm el}$ of the system.
\subsection{The second quantisation of the electronic Hamiltonian}
For the electronic Hamiltonian and the interaction with the electromagnetic field we introduce the formalism of second quantisation, a standard technique of quantum chemistry \cite{FeWa71,SzOs96} that is also frequently used in the treatment of x-ray matter interaction \cite{veenendaal}. In this representation, similar to the representation of the quantised electromagnetic field, electrons are created by appropriate field operators that act on the vacuum state. As in the theory of the quantised electromagnetic field, different representations of the field operators can be chosen, that rely on the expansion in terms of modes or complete single-particle basis sets. Here we choose
a representation that underpins on the mode expansion of coordinate space by delta functions, so that the field operators are given by the two-component spinor 
\begin{equation}
\hat{\psi}({\bm x}) = \left(\begin{array}{c} \hat{\psi}_{+1/2}({\bm x}) \\ \hat{\psi}_{-1/2}({\bm x})\end{array}\right).
\end{equation}
and its Hermitian conjugate ($\hat{\psi}^{\dag}({\bm x})$). In that representation $\hat{\psi}_{\sigma}({\bm x})$ and $\hat{\psi}^{\dag}_{\sigma}({\bm x})$ annihilates and respectively creates an electron of spin $\sigma$ at position $\bm x$. The fermionic creation and annihilation operators fulfil the usual fermionic anti-commutation relations:
\begin{equation}
\{\hat{\psi}_{\sigma}({\bm x}),\hat{\psi}_{\sigma'}({\bm x}')\}=
\{\hat{\psi}^{\dag}_{\sigma}({\bm x}),\hat{\psi}^{\dag}_{\sigma'}({\bm x}')\} = 0,\;
\{\hat{\psi}_{\sigma}({\bm x}),\hat{\psi}^{\dag}_{\sigma'}({\bm x}')\} = \delta_{\sigma,\sigma'} \delta^{(3)}({\bm x} - {\bm x}')\;.
\end{equation}
In this representation the electronic Hamiltonian of eq. (\ref{hel}) is rewritten as
\begin{eqnarray}\label{helquantised}
\hat{H}_{{\mathrm el}}& =& 
\int {\rm d}^3{\bm x}\hat{\psi}^{\dag}({\bm x})\left\{-\frac{1}{2}\nabla^2 - \sum_n\frac{Z_n}{|{\bm x} -{\bm R}_n|}\right\}\hat{\psi}({\bm x})\nonumber\\
&& + \frac{1}{2} \int {\rm d}^3{\bm x} \int {\rm d}^3{\bm x}' \;
\frac{\hat{\psi}^{\dag}({\bm x})\hat{\psi}^{\dag}({\bm x}')\hat{\psi}({\bm x}')\hat{\psi}({\bm x})}{|{\bm x} - {\bm x}'|}\;.
\end{eqnarray}
The advantage of the representation of the Hamiltonian in second quantisation is, that it is independent of the number of electrons of the system. The number of electrons, in analogy to the number of photons in a Fock states of the electromagnetic field, is fixed by the quantum state of the electrons.  Likewise, other observables and operators can be represented in second quantisation. A useful expressions that we will encounter in the following tutorial is for example the electronic density operator
\begin{equation}
\hat{n}({\bm x})=\hat{\psi}^\dag({\bm x}) \hat{\psi}({\bm x}).
\end{equation}
A general one-body operator that is given in space representation by the function $j({\bm x})$, can be expressed within the second quantisation by
\begin{equation}
\hat{j}=\int {\rm d}^3{\bm x}\;\hat{\psi}^\dag({\bm x}) j({\bm x}) \hat{\psi}({\bm x})\;.
\end{equation}
Accordingly, the matter-field interaction Hamiltonian eq.\ (\ref{hint0}) resulting from the minimal coupling Hamiltonian can be expressed in second quantisation and reads
\begin{equation}
\label{hint}
\hat{H}_{\mathrm{int}}  =  \alpha\!\! \int {\rm d}^3{\bm x} \hat{\psi}^{\dag}({\bm x}) 
\left[\hat{{\bm A}}({\bm x})\cdot\frac{{\bm \nabla}}{\rm i}\right] \hat{\psi}({\bm x})
 + \frac{\alpha^2}{2} \int {\rm d}^3{\bm x}\hat{\psi}^{\dag}({\bm x}) \hat{\bm A}^2({\bm x}) \hat{\psi}({\bm x}).
\end{equation}
The first term of eq.\ (\ref{hint}) corresponds to the $\hat{p}\cdot\hat{A}$-interaction, linear in the vector potential, and as we shall see, gives rise to photoabsorption, resonance scattering and anomalous diffraction. The second interaction, quadratic in the vector potential (the $\hat{A}^2$-interaction) is usually neglected for visible wavelengths. In the x-ray region, the $\hat{A}^2$ contribution gives rise to elastic and incoherent Thomson scattering. \\
\indent A different way of representing the electronic Hamiltonian in second quantisation relies on the expansion of the electronic wave functions in appropriate single-particle orbitals. Let us suppose that the correlated electronic wave function of the considered molecular system is expanded in appropriate (but in principle arbitrary) single-particle spin orbitals $|\varphi_p\rangle$, where $p$ stands for a complete list of quantum numbers characterising the eigenstate (including the spin degree of freedom), that are eigenfunctions of a one-particle Hamiltonian $\hat{F}$ (for example a mean-field Hamiltontian, the Hartree-Fock Hamiltonian, the Kohn-Sham Hamiltonian, or similar):
\begin{equation}
\label{eq45a}
\hat{F}|\varphi_p\rangle = \varepsilon_p |\varphi_p\rangle,
\end{equation}
where $\varepsilon_p$ is the orbital energy. 
We then define creation $\hat{c}^\dag_p$ and annihilation operators $\hat{c}_p$ in the formulation of second quantisation, fulfilling the fermionic anti-commutation relations
\begin{equation}
\{\hat{c}_p,\hat{c}_{p'}\}=
\{\hat{c}^{\dag}_p,\hat{c}^{\dag}_{p'}\}= 0\;\;
\mathrm{and}\;\;
\{\hat{c}_p,\hat{c}^{\dag}_{p'}\} = \delta_{p,p'}\;.
\end{equation}
The Hamiltonian $\hat{F}$ then has the expansion
\begin{equation}
\label{eq48a}
\hat{F} = \sum_p \varepsilon_p \hat{c}_p^{\dag} \hat{c}_p.  
\end{equation}
To switch between the two representations in second quantisation, the field operators $\hat{\psi}({\bm x})$ and $\hat{\psi}^{\dag}({\bm x})$ can be expanded in the orbital basis as follows \cite{FeWa71}:
\begin{equation}\label{psiexp}
\hat{\psi}({\bm x}) = \sum_p \varphi_p({\bm x}) \hat{c}_p\;\;\mathrm{and}\;\;
\hat{\psi}^{\dag}({\bm x}) = \sum_p \varphi_p^{\dag}({\bm x}) \hat{c}^{\dag}_p.
\end{equation}
An electronic state can be created by the action of the creation operators on the vacuum state $|\mathrm{vacuum}\rangle$. For instance, the $N_{\mathrm{el}}$-electron ground-state of the system can be build up by:
\begin{equation}
\label{HFgroundstate}
|\Psi_0^{N_{\mathrm{el}}}\rangle  \approx  |\Phi_0^{N_{\mathrm{el}}}\rangle
 \equiv \prod_{i=1}^{N_{\mathrm{el}}}\hat{c}_i^{\dag} |\mathrm{vacuum}\rangle, \nonumber
\end{equation}
where we suppose that the many-body ground-state $|\Psi_0^{N_{\mathrm{el}}}\rangle$ can be approximated by a Slater determinant $|\Phi_0^{N_{\mathrm{el}}}\rangle$ with occupation of the $N_{\mathrm{el}}$ lowest orbitals. In the following we suppose that the initial state of the system is its electronic ground state.

\section{Perturbation theory in the interaction picture}
We treat the matter-field interaction in the limit of perturbation theory. To that end, we partition the Hamiltonian in $\hat{H}=\hat{H}_0+\hat{H}_{\mathrm{int}}$, where the unperturbed Hamiltonian $\hat{H}_0= \hat{H}_{\mathrm{mol}} + \hat{H}_{\mathrm{EM}}$ is given by the sum of the Hamiltonian of the electromagnetic field and the molecular Hamiltonian  and the matter-field interaction Hamiltonian $\hat{H}_{\mathrm{int}}$ is given by Eq.\ \ref{hint}. We employ the interaction picture. The state vector in the interaction picture is given by
\begin{equation}
\label{eq35}
|\Psi,t\rangle_{\mathrm{int}} = {\rm e}^{{\rm i} \hat{H}_0 t} |\Psi,t\rangle
\end{equation}
and we suppose that the system starts from an initial state $|I\rangle$, with the initial condition $\lim_{t\rightarrow -\infty}|\Psi,t \rangle_{\mathrm{int}}=\lim_{t\rightarrow -\infty}|\Psi,t\rangle=|I\rangle$.
The Schr\"odinger equation in the interaction picture is then
\begin{equation}
\label{eq36}
{\rm i} \frac{\partial}{\partial t}|\Psi,t\rangle_{\mathrm{int}} = \lim_{\epsilon\rightarrow +0}{\rm e}^{{\rm i} \hat{H}_0 t} \hat{H}_{\mathrm{int}} {\rm e}^{-\epsilon |t|}
{\rm e}^{-{\rm i} \hat{H}_0 t} |\Psi,t\rangle_{\mathrm{int}}.
\end{equation} 
Here, we adopt the usual adiabatic switching of the interaction by the exponential term ${\rm e}^{-\epsilon |t|}$. The formal integration of eq.\ \ref{eq36} results in the
 perturbative expansion of the state vector in interaction picture
\begin{eqnarray}
\label{expansion}
|\Psi,t\rangle_{\mathrm{int}} & = & |I\rangle -{\rm i} \int_{-\infty}^t {\rm d}t' {\rm e}^{{\rm i} \hat{H}_0 t'} \hat{H}_{\mathrm{int}} {\rm e}^{-\epsilon |t'|}
{\rm e}^{-{\rm i} \hat{H}_0 t'}|I\rangle\nonumber \\
& & -\int_{-\infty}^t {\rm d}t' {\rm e}^{{\rm i} \hat{H}_0 t'} \hat{H}_{\mathrm{int}} {\rm e}^{-\epsilon |t'|} 
{\rm e}^{-{\rm i} \hat{H}_0 t'}
 \times \int_{-\infty}^{t'} {\rm d}t'' {\rm e}^{{\rm i} \hat{H}_0 t''} \hat{H}_{\mathrm{int}} {\rm e}^{-\epsilon |t''|} {\rm e}^{-{\rm i} \hat{H}_0 t''}|I\rangle + \ldots \;.
\end{eqnarray}
This perturbative expansion along with the structure of the interaction Hamiltonian of eq.\ (\ref{hint}) gives an expansion of the interaction in increasing order of the $p\cdot\hat{A}$ and $\hat{A}^2$ interaction terms.\\
\indent The transition amplitude from the forward propagated initial state $|I\rangle$ to an arbitrarily chosen final state $|F\rangle$ in the limit of large times defines the S-matrix of the interaction. In the following we will treat up to second order interactions with the photon field. The transition amplitude up to second order in the interaction Hamiltonian then reads:
\begin{eqnarray}
\label{Smatrix}
S_{FI}
& = & \lim_{t\rightarrow\infty}\langle F|\Psi,t\rangle_{\mathrm{int}}
=-{\rm i} \int_{-\infty}^{\infty} {\rm d}t {\rm e}^{{\rm i} (E_F-E_I) t} \langle F|\hat{H}_{\mathrm{int}}|I\rangle \nonumber\\
&& -\int_{-\infty}^{\infty} {\rm d}t\; 
\sum_{M} {\rm e}^{{\rm i} (E_F-E_M)t} \langle F|\hat{H}_{\mathrm{int}}|M\rangle \nonumber
 \times \int_{-\infty}^{t} {\rm d}t' \;{\rm e}^{{\rm i} (E_M-E_I -{\rm i} \epsilon) t'} 
 {\rm e}^{-\epsilon |t'|}
 \langle M|\hat{H}_{\mathrm{int}}|I\rangle
 \nonumber\\
&=& -2\pi {\rm i} \; \delta(E_F-E_I)\;\langle F|\hat{H}_{\mathrm{int}}|I\rangle  \nonumber\\
&& -\int_{-\infty}^{\infty} {\rm d}t \sum_{M} {\rm e}^{{\rm i} (E_F-E_M)t} 
\langle F|\hat{H}_{\mathrm{int}}|M\rangle
\times {\rm e}^{{\rm i} (E_M-E_I) t} \frac{\langle M|\hat{H}_{{\rm int}}|I\rangle}{{\rm i} (E_M-E_I -\rm i \epsilon)} 
\nonumber\\
&=& -2\pi {\rm i}\; \delta(E_F-E_I)\;
\left\{\langle F|\hat{H}_{\mathrm{int}}|I\rangle + \sum_{M}\frac{\langle F |\hat{H}_{\mathrm{int}}|M\rangle
\langle M|\hat{H}_{\mathrm{int}}|I\rangle}{E_I-E_M + {\rm i} \epsilon}\right\}, 
\end{eqnarray}
where we set $\epsilon$ to zero at the appropriate places.
The transition probability is given by $|S_{FI}|^2$ and would involve the square of the energy-conserving delta function. Expressing the square of the delta function by
\begin{equation}
[\delta(E_F-E_I)]^2  = \lim_{T\rightarrow \infty} \delta(E_F-E_I) \int_{-T/2}^{T/2} \frac{{\rm d}t}{2\pi} {\rm e}^{{\rm i} (E_F-E_I) t} \\
 = \lim_{T\rightarrow \infty} \delta(E_F-E_I) \frac{T}{2\pi}\;
\end{equation}
for the limit of large $T$, one then defines a transition rate $\Gamma_{FI}:=\frac{|S_{FI}|^2}{T}$, that in second order is given by
\begin{equation}\label{gamma}
  \Gamma_{FI}= 2\pi \delta(E_F-E_I)\left|
\langle F|\hat{H}_{\mathrm{int}}|I\rangle
+ \sum_{M}\frac{\langle F |\hat{H}_{\mathrm{int}}|M\rangle
\langle M|\hat{H}_{\mathrm{int}}|I\rangle}{E_I-E_M + {\rm i} \epsilon}\right|^2. 
\end{equation}
This expression for the transition rate will be the basis for deriving cross sections of different x-ray matter interaction processes. To that end, the initial and final states of the combined matter-radiation system have to be defined and all possible interaction terms have to be picked, that connect initial and final states. The coherent sum of those pathways then will give a transition rate. 
In the following chapter, we shall show details of this general strategy by calculating the process of x-ray absorption, the derivation of more complicated processes involving second order terms follows along similar lines, but is slightly more tedious.
\section{X-ray absorption}
One of the strongest effects of x-ray matter interaction is the absorption of an x-ray photon, which shall be derived here in the limit of first order perturbation theory. We shall start out our analysis with a molecule in its many-body ground state $|\Psi_0^{N_{\rm{el}}}\rangle$. We suppose that the incoming x-ray field is in a Fock state with $N_{\rm{EM}}$ photons occupying a single mode of a chosen momentum and polarisation $(\bm{k_{\rm in}},\lambda_{\rm in})$. The combined initial state of molecule plus field is hence
\begin{equation}\label{initialstate}
    |I\rangle=|\Psi_0^{N_{\rm{el}}}\rangle\otimes|N_{\rm EM}\rangle.
\end{equation}
Absorption of a photon then results in a Fock state with one less photon and a general molecular final state $|\Psi_F^{N_{\rm{el}}}\rangle\neq|\Phi_0^{N_{\rm{el}}}\rangle$:
\begin{equation}
    |F\rangle=|\Psi_F^{N_{\rm{el}}}\rangle\otimes|N_{\rm EM-1}\rangle\;.
\end{equation}
We calculate the transition rate from the initial to final state according to eq.\ (\ref{gamma}). Keeping only terms linear in the interaction with the vector potential then gives
\begin{eqnarray}\label{gabs1}
 \Gamma_{FI} &=& 2\pi\delta(E_F-E_I)\alpha^2\nonumber\\
&& \left|\langle \Psi_F^{N_{\rm{el}}}|\otimes\langle N_{\rm EM-1}|
\int {\rm d}^3x \;
\hat{\psi}^{\dag}({\bm x}) 
 \left[\hat{{\bm A}}({\bm x})\cdot\frac{{\bm \nabla}}{\rm i}\right] \hat{\psi}({\bm x})
|N_{\rm EM}\rangle\otimes|\Psi_0^{N_{\rm{el}}}\rangle
\right|^2. 
\end{eqnarray}
Now, we make use of the expansion of the vector potential of eq.\ (\ref{vectorpotential}). With the restriction of the incoming and outgoing state vector of the electromagnetic field, only the term proportional to $\hat{a}_{{\bm k_{\rm in}},\lambda_{\rm in}}$ of the expansion of the vector potential gives a contribution non equal to zero, so that eq.\ (\ref{gabs1}) becomes
\begin{eqnarray}\label{gabs2}
 \Gamma_{FI} &=& (2\pi)^2\delta(E_F-E_I) \frac{1}{V \omega_{\rm{in}} }\nonumber\\
&\times& \left|\int {\rm d}^3{\bm x}\;
\langle \Psi_F^{N_{\rm{el}}}|\hat{\psi}^{\dag}({\bm x}) 
{\rm e}^{{\rm i}{\bm k_{\rm in}}\cdot{\bm x}}
{\bm \epsilon}_{{\bm k_{\rm in}},\lambda_{\rm in}}\cdot\frac{{\bm \nabla}}{\rm i} \hat{\psi}({\bm x})
|\Psi_0^{N_{\rm{el}}}\rangle
\langle N_{\rm EM-1}|\hat{a}_{{\bm k},\lambda}|N_{\rm EM}\rangle
\right|^2. 
\end{eqnarray}
The photon annihilation operator acting on the Fock state gives 
\begin{equation}
\hat{a}_{{\bm k_{\rm in}},\lambda_{\rm in}}|N_{\rm EM}\rangle=\sqrt{N_{\rm EM}}|N_{\rm EM}-1\rangle\;.\nonumber
\end{equation}
The energy of the initial and final state are given by $E_I=E_0^{N_{\rm el}}+\omega_{\rm{in}}N_{\rm EM} $ and  $E_F=E_F^{N_{\rm el}}+\omega_{\rm{in}}(N_{\rm EM}-1)$ respectively. 
The incoming photon flux is given by 
\begin{equation}\label{flux}
J_{\rm EM} =\frac{1}{\alpha}\frac{N_{\rm EM}}{V}\;.
\end{equation}
We therefore can define a cross section for photoabsorption to the specific final electronic state $|\Psi_F^{N_{\rm{el}}}\rangle$ by
\begin{equation}
    \sigma_F({\bm k}_{\mathrm{in}},\lambda_{\mathrm{in}})=\Gamma_{FI} / J_{\rm EM}
\end{equation}
and we get
\begin{eqnarray}\label{gabs3}
 \sigma_F({\bm k}_{\mathrm{in}},\lambda_{\mathrm{in}})
 &= &
 4\pi^2\delta(E_F^{N_{\rm el}}-E_0^{N_{\rm el}} -\omega_{\rm{in}}) \frac{\alpha}{\omega_{\rm{in}} } 
 \nonumber\\&& \times 
 \left|\int {\rm d}^3{\bm x}\;
\langle \Psi_F^{N_{\rm{el}}}|\hat{\psi}^{\dag}({\bm x}) 
{\rm e}^{{\rm i}{\bm k_{\rm in}}\cdot{\bm x}}
{\bm \epsilon}_{{\bm k_{\rm in}},\lambda_{\rm in}}\cdot\frac{{\bm \nabla}}{\rm i} \hat{\psi}({\bm x})
|\Psi_0^{N_{\rm{el}}}\rangle
\right|^2. 
\end{eqnarray}
This is the general expression for the photoabsorption cross section to an electronic state $|\Psi_F^{N_{\rm{el}}}\rangle$. Aiming for a more intuitive interpretation, we now express the field creation and annihilation operators in an expansion of appropriate single-particle orbitals. With eq.\ (\ref{psiexp}) we get 
\begin{eqnarray}\label{sabsex}
 \sigma_F({\bm k}_{\mathrm{in}},\lambda_{\mathrm{in}})&=& 4\pi^2\delta(E_F^{N_{\rm el}}-E_0^{N_{\rm el}} -\omega_{\rm{in}}) \frac{\alpha}{\omega_{\rm{in}} }
  \nonumber\\
&&
 \times 
\left|\sum_{p,a}
 \underbrace{\left[\int {\rm d}^3{\bm x}\;
 \varphi^*_p({\bm x})
 {\rm e}^{{\rm i}{\bm k_{\rm in}}\cdot{\bm x}}
{\bm \epsilon}_{{\bm k_{\rm in}},\lambda_{\rm in}}\cdot\frac{{\bm \nabla}}{\rm i} 
 \varphi_a({\bm x})\right]}
 _{\langle \varphi_p|
 {\rm e}^{{\rm i}{\bm k_{\rm in}}\cdot{\bm x}}
{\bm \epsilon}_{{\bm k_{\rm in}},\lambda_{\rm in}}\cdot\frac{{\bm \nabla}}{\rm i} 
 |\varphi_a\rangle}
\times
\underbrace{
\langle\Psi_F^{N_{\rm{el}}}| \hat{c}^\dag_p \hat{c}_a|\Psi_0^{N_{\rm{el}}}\rangle}_{\rm overlap\; integral}
\right|^2. 
\end{eqnarray}
The term in brackets is the single-particle transition matrix element from orbital $\varphi_a$ to $\varphi_p$, that can be rewritten in the usual bra-ket notation. The second term following the transition matrix is a kind of overlap integral that, starting from the initial state vector, describes the annihilation of an electron in spin-orbital $\varphi_a$ and the creation of an electron in orbital $\varphi_p$ by the process of photo absorption. This overlap integral is still exact in terms of the electron-electron interaction, since electron correlations are in principle contained in the electronic state vectors $|\Psi_0^{N_{\rm{el}}}\rangle$ and $|\Psi_F^{N_{\rm{el}}}\rangle$. 
\\
\indent To reduce the complexity, we now adopt the mean-field approximation. We replace the many-body Hamiltonian $\hat{H}_{\rm el}$ of eq.\ (\ref{helquantised}) by the Fock operator $\hat{F}$ of eq.\ (\ref{eq48a}).
The ground-state $|\Psi_0^{N_{\rm{el}}}\rangle$ of the system is then approximated by the Slater determinant $|\Phi_0^{N_{\rm{el}}}\rangle$ given in eq.\ (\ref{HFgroundstate}). The ground-state energy is then approximated by the sum of single-particle energies $E_0^{N_{\rm el}}\approx\sum_{i=1}^{N_{\rm el}}\varepsilon_i $. The overlap integral between initial and final state, which was still exact in eq.\ (\ref{sabsex}) then reduces to
\begin{eqnarray}
\langle\Psi_F^{N_{\rm{el}}}| \hat{c}^\dag_p \hat{c}_a|\Psi_0^{N_{\rm{el}}}\rangle\approx
\langle\Psi_F^{N_{\rm{el}}}|
\underbrace{\hat{c}^\dag_p \hat{c}_a
\Pi_{i=1}^{N_{\rm el}}\hat{c}^\dag_i|{\rm vacuum}\rangle}
_{:=|\Phi_a^p\rangle}\;,
\end{eqnarray}
where we defined the state $|\Phi_a^p\rangle$, which denotes a single particle-hole excited state of the ground state, with a hole in an initially occupied spin orbital $\varphi_a$ and an excitation in spin-orbital $\varphi_p$. The approximated overlap integral hence reduces the sum in Eq.\ (\ref{sabsex}) to spin orbitals $a$, that are initially occupied, and spin orbitals $p$ that are initially unoccupied in the ground state. Although the final state $|\Psi_F^{N_{\rm{el}}}\rangle$ is general, within the mean-field approximation one can fix it by $|\Phi_a^p\rangle$, so that the overlap integral is unity. The orbital $p$ can be a bound or continuum orbital, describing resonant excitation or ionisation, respectively.\\
\begin{figure}
    \centering
    \includegraphics[width=11cm]{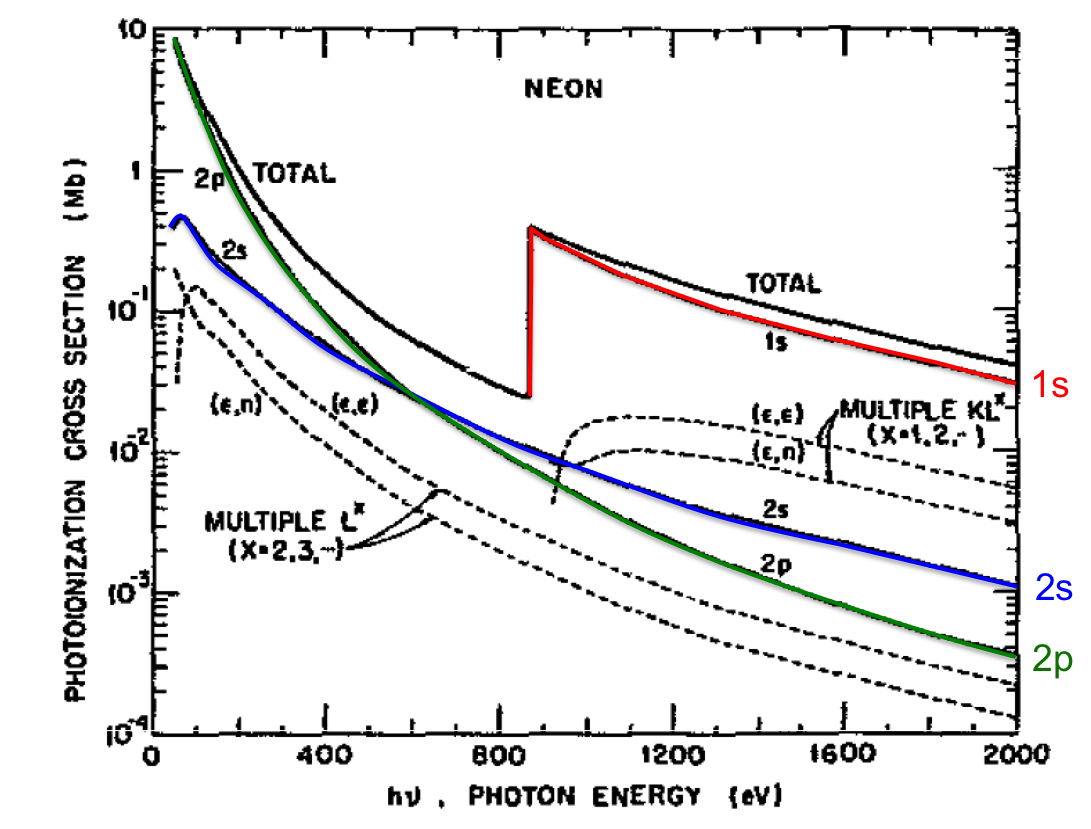}
    \caption{Calculated sub-shell and total photoionisation cross section for Neon reproduced from Ref. \cite{Krause}.}
    \label{fig:1}
\end{figure}
We now suppose that the incoming photon energy $\omega_{\rm{in}}$ exceeds the ionisation potential $I_a=-\epsilon_a$ of state $\varphi_a$. In this case, we then can define an ionisation cross section that promotes an electron from orbital $a$ to orbital $p$ as
\begin{eqnarray}\label{subshell}
 \sigma_a^p({\bm k}_{\mathrm{in}},\lambda_{\mathrm{in}})&=& \delta(\varepsilon_p+I_a -\omega_{\rm{in}}) \frac{4\pi^2\alpha}{\omega_{\rm{in}} }
\left|
 \langle \varphi_p|
 {\rm e}^{{\rm i}{\bm k_{\rm in}}\cdot{\bm x}}
{\bm \epsilon}_{{\bm k_{\rm in}},\lambda_{\rm in}}\cdot\frac{{\bm \nabla}}{\rm i} 
 |\varphi_a\rangle
\right|^2. 
\end{eqnarray}
The sub-shell ionisation cross section from orbital $a$ can then be defined by an integral over all possible final states
\begin{equation}
    \sigma_a({\bm k}_{\mathrm{in}},\lambda_{\mathrm{in}})=\int_p \!\!\!\!\!\!\!\!\!\sum \sigma_a^p({\bm k}_{\mathrm{in}},\lambda_{\mathrm{in}})
\end{equation}
\begin{figure}
    \centering
    \includegraphics[width=8cm]{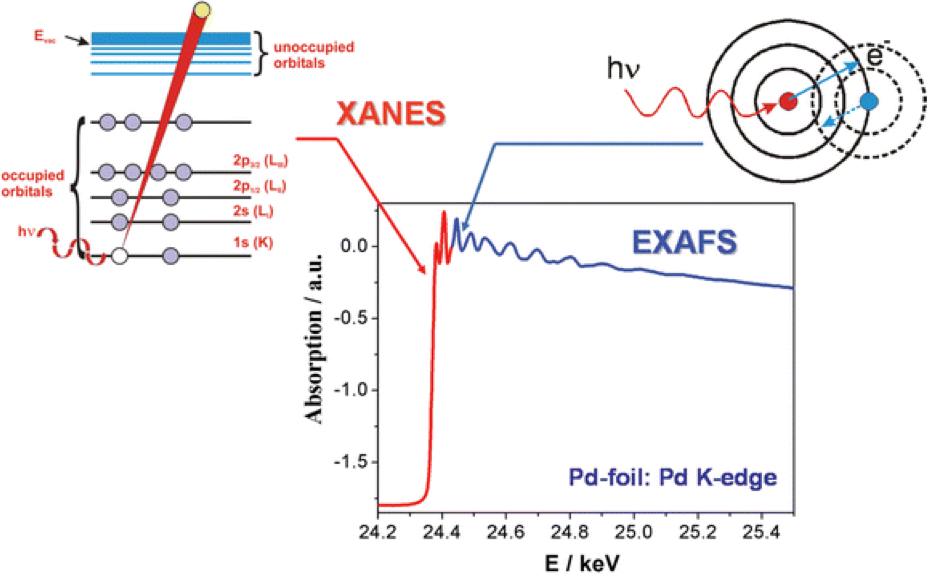}
    \caption{Photoabsorption Cross Section of a Pd foil. Principle of X-ray absorption near edge structure (XANES) and extended X-ray absorption fine structure (EXAFS) spectroscopy, reproduced from \cite{Grunwaldt}.}
    \label{fig:2}
\end{figure}
As an example, Fig.\ (\ref{fig:1}) shows calculated sub-shell, and total photoionisation cross sections for Neon from Ref. \cite{Krause} in a mean-field approximation that is based on a parameterised Hartree-Fock-Slater potential. The 1s photoionisation cross section right at the ionisation edge is of the order of 1 Mbarn and falls off rapidly and monotonously  for energies above the 1s absorption edge. Neglecting doubly-excited states in the continuum, the cross section does not show any particular structure for energies above the ionisation edge. In molecules and solids, on the contrary, the cross sections above the edge show characteristic structures (see Fig.\ \ref{fig:2}) that in principle contain structural information of the multi-atomic system. There are two experimental methods that rely on this structure, namely {\it X-ray Absorption Near Edge Structure} (XANES)  that looks into structure of the photoionisation cross section right after the sub-shell ionisation edge and {\it Extended X-ray Absorption Fine Structure} (EXAFS), that looks at structures at the order of 50-100 eV above the edge. These characteristic structures contain information of many-body effects that are contained in eq.\ (\ref{sabsex}). The XANES structure \cite{Henderson} is mainly dominated by the overlap integral of eq.\ (\ref{sabsex}). In practice, XANES structures of metals are often dominated by so called Multiplet effects in relation to spin-orbit splitting. Thus, on the level of theory we have presented here, these effects are beyond our theoretical description. Another contribution for the XANES structure is often described in the term of "core-hole effects", meaning that the final state cannot be expressed in a single-particle hole excitation based on the single-particle orbitals of the ground state, but in general shows configuration mixing, that results out of the reshaping of the spin orbitals in the final state due to the presence of a core hole. In our approach this would mean, that the electronic final state in eq.\ (\ref{sabsex}) would need to account for electron correlations.  For the theory of XANES spectroscopy see, for example Vol.\ 1 chapt.\ 4 of reference \cite{bokhoven1}. \\
\indent The physical interpretation the EXAFS structures \cite{Lee} relies on the picture of scattering of the outgoing photo-electron on neighbouring atomic cores. A photo-electron of 50 eV has a De Broglie wavelength of roughly 1.7  \AA, a typical size of internuclear distances in molecular systems and solids. The EXAFS method is hence highly sensitive to changes in the bond length. More information on EXAFS can be found for example in Vol.\ 1 chapt.\ 11 of reference \cite{bokhoven1}.

\section{Elastic X-ray scattering}
Elastic or coherent x-ray scattering describes a photon in / photon out process for which it is assumed that incoming and outgoing photon have the same energy, but can have different momenta. The material system remains unchanged in the elastic scattering process. Starting from an initial state given by eq.\ (\ref{initialstate}), that supposes the system to be initially in the electronic ground state and supposing a single-mode of the incoming field, the final state of coherent x-ray scattering has thus the general form
\begin{equation}\label{finalelastic}
    |F\rangle = 
    |\Psi_0^{N_{\rm{el}}}\rangle \hat{a}^\dag_{{\bm k}_{F},\lambda_F}  
    |N_{\mathrm{EM}}-1\rangle\; 
    \mbox {, with } =|{\bm k}_F|=|{\bm k}_{\mathrm{in}}|.
\end{equation}
A photon in / photon out process can be mediated by two contributions of the interaction Hamiltonian eq.\ (\ref{hint}): within first order perturbation theory, the $\hat{A}^2$ induces a quasi simultaneous absorption and emission of the photon ("contracted vertex"). This is the scattering contribution dominating scattering for photon energies far above core-ionisation edges of the system. For photon energies close to ionisation edges or resonances, the $\hat{p\cdot\hat{A}}$ term in second order perturbation theory has to be taken into account, to accurately derive the elastic scattering cross section. This term gives dispersion corrections to the scattering process and its strong dependence on the photon energy is used in an experimental method that is called {\it Multi-wavelength Anomalous Diffraction (MAD)}, to solve the so-called phase problem of elastic x-ray scattering. \\
\indent According to eq.\ (\ref{gamma}), the general expression for the transition rate from the incoming state of eq.\ (\ref{initialstate}) to the final state of structure eq.\ (\ref{finalelastic}) is, up to second order contributions in $\hat{A}$ given by 
\\
\\
\begin{eqnarray}
\label{elastic}
\Gamma_{FI} & = & 
2\pi \delta(\omega_F -\omega_{\rm{in}})\alpha^4\nonumber\\
&\times&\left|
\frac{1}{2} \int {\rm d}^3{\bm x}\;
\langle\Psi_0^{N_{\rm{el}}}|\langle N_{\mathrm{EM}}-1|\hat{a}_{{\bm k}_{F},\lambda_{F}}
\hat{\psi}^{\dag}({\bm x})
\hat{A}^2({\bm x})\hat{\psi}({\bm x})|\Psi_0^{N_{\mathrm{el}}}\rangle|N_{\mathrm{EM}}\rangle \right . \nonumber\\
&+& \int {\rm d}^3{\bm x}\int {\rm d}^3{\bm x}'
\sum_{M} \left[\langle\Psi_0^{N_{\rm{el}}}|\langle N_{\mathrm{EM}}-1|\hat{a}_{{\bm k}_{F},\lambda_{F}}  \hat{\psi}^{\dag}({\bm x})\hat{A}({\bm x})\cdot\frac{\nabla}{i}\hat{\psi}({\bm x})\right.\nonumber\\
&\times&
\underbrace{
\frac{| \Psi_\mathrm{M}\rangle |N_{\mathrm{EM}}-1\rangle 
\langle N_\mathrm{EM}-1| \langle \Psi_\mathrm{M}| } 
{       E^{N_{\rm el}}_0 - E^{N_{\rm el}}_M+\omega_{\rm{in}}
 + {\rm i} \varepsilon}}_{\mbox{regular pathway}}\left.
\hat{\psi}^{\dag}({\bm x'})\hat{A}({\bm x'})\cdot\frac{\nabla}{i}
\hat{\psi}({\bm x'})
|\Psi_0^{N_{\mathrm{el}}}\rangle|N_{\mathrm{EM}}\rangle\right]\nonumber\\
&+& \int {\rm d}^3{\bm x}\int {\rm d}^3{\bm x}'
\sum_{M,m}\left[ \langle\Psi_0^{N_{\rm{el}}} |\langle N_{\mathrm{EM}}-1|\hat{a}_{{\bm k}_{F},\lambda_{F}}  \hat{\psi}^{\dag}({\bm x})\hat{A}({\bm x})\cdot\frac{\nabla}{i}\hat{\psi}({\bm x})\right.\nonumber\\
&\times&\underbrace{
\frac{ |\Psi_\mathrm{M}\rangle \hat{a}^\dag_{{\bm k}_{m},\lambda_{m}}| N_{\mathrm{EM}}\rangle 
\langle N_\mathrm{EM}|\hat{a}_{{\bm k}_{m},\lambda_{m}}| \langle \Psi_\mathrm{M}| } 
{E^{N_{\rm el}}_0 - E^{N_{\rm el}}_M-\omega_{m}
+{\rm i} \varepsilon}}_{\mbox{irregular pathway}}
\left.\left.
 \hat{\psi}^{\dag}({\bm x'})\hat{A}({\bm x'})\cdot\frac{\nabla}{i}
\hat{\psi}({\bm x'})
|\Psi_0^{N_{\mathrm{el}}}\rangle|N_{\mathrm{EM}\rangle}\right]
\right|^2
\end{eqnarray}
The second order $\hat{p}\cdot\hat{A}$ contribution to the cross section has two terms, one corresponding to the regular pathway, for which an incoming photon is absorbed, followed by the emission of a photon with the outgoing wave vector ${\bm k}_F$. The second contribution is a counter-intuitive pathway, where emission of the scattered photon happens before annihilation of a photon in the incoming mode. This is sometimes also referred to as time-inverted pathway. The following analysis and derivation will be restricted to the coherent $\hat{A}^2$ contribution of scattering, assuming that the photon energies are far above any resonances and ionisation edges. The contribution due to dispersion corrections close to resonances and ionisation edges are discussed in Ref. \cite{santra}.
Inserting the quantisation of the vector potential of eq.\ \ref{vectorpotential} into eq.\ \ref{elastic} and only keeping those therms that lead to the specific structure of the final state we get
\begin{eqnarray}
\label{elasticThomson}
\Gamma_{FI}  &=&  
2\pi \delta(\omega_F -\omega_{\rm{in}})\frac{\alpha^4}{4}\nonumber\\
&&\times \left|\int {\rm d}^3{\bm x}\;  
 \langle\Psi_0^{N_{\mathrm{el}}}|\langle N_{\mathrm{EM}}-1|\hat{a}_{{\bm k}_{F},\lambda_{F}}
\hat{\psi}^{\dag}({\bm x}) \hat{A}^2({\bm x})\hat{\psi}({\bm x})
|\Psi_0^{N_{\mathrm{el}}}\rangle|N_{\mathrm{EM}}\rangle \right |^2\nonumber\\
&=&
\frac{(2\pi)^3}{V}\frac{N_{EM}}{V}\frac{1}{ \omega_{\rm{in}}^2} 
\delta(\omega_F -\omega_{\rm{in}})
\left|{\bm \epsilon}^{\ast}_{{\bm k}_F,\lambda_F} \cdot {\bm \epsilon}_{{\bm k}_{\mathrm{in}},\lambda_{\mathrm{in}}} \right|^2\nonumber\\
&&\times\left|\int {\rm d}^3{\bm x}\; \langle\Psi_0^{N_{\mathrm{el}}}|\hat{\psi}^{\dag}({\bm x}){\rm e}^{{\rm i}({\bm k}_{\mathrm{in}}-{\bm k}_F)\cdot{\bm x}}
\hat{\psi}({\bm x})|\Psi_0^{N_{\mathrm{el}}}\rangle\right|^2\;.
\end{eqnarray}
We define the photon momentum transfer $\bm Q=\bm k_{in}- \bm k_F$. The expression in the second line of eq.\ \ref{elasticThomson} can be simplified to
\begin{eqnarray}
\int {\rm d}^3{\bm x} \;\langle\Psi_0^{N_{\mathrm{el}}}|\hat{\psi}^{\dag}({\bm x})
{\rm e}^{{\rm i}{\bm Q}\cdot{\bm x}}
\hat{\psi}({\bm x})|\Psi_0^{N_{\mathrm{el}}}\rangle&=&
\int {\rm d}^3{\bm x}\;
{\rm e}^{{\rm i}{\bm Q}\cdot{\bm x}}
\langle\Psi_0^{N_{\mathrm{el}}}|\hat{\psi}^{\dag}({\bm x})
\hat{\psi}({\bm x})|\Psi_0^{N_{\mathrm{el}}}\rangle\nonumber\\
&=&\underbrace{\int {\rm d}^3{\bm x}\;
{\rm e}^{{\rm i}{\bm Q}\cdot{\bm x}}\rho_0({\bm x})}_
{=:f^0({\bm Q})}
\;.
\end{eqnarray}
Here, we made use of the fact, that $\hat{n}({\bm x})=\hat{\psi}^{\dag}({\bm x})\hat{\psi}({\bm x})$ stands for the density operator and we defined the ground-state density as $\rho_0({\bm x})$.
The Fourier transform of the ground-state density is defined as the elastic scattering factor, or elastic form factor $f^0({\bm Q})$.
\\
In a typical coherent scattering experiment, the outgoing photon flux is measured in a certain direction ${\bm k}_F$ within solid angle $\Delta\Omega$. Generally the polarisation and the energy of the outgoing photon are not measured. We define the partial differential scattering cross section of the elastic scattering process by
\begin{eqnarray}
{\rm d}\sigma ({\bm Q})=
\frac{1}{J_{\rm EM}}
\sum_{\lambda_{\mathrm{F}}, \bm \Delta\Omega,\omega_F}\Gamma_{FI} \;.
\end{eqnarray}
The sum can be rewritten in terms of an integral
\begin{equation}
\sum_{{\bm \Delta\Omega}, \omega_F} =\alpha^3 \frac{V}{(2\pi)^3}{\rm d}{\bm \Omega} \int_0^\infty\omega_F^2\;{\rm d}\omega_F
\end{equation}
so that we finally get
\begin{eqnarray}
\frac{{\rm d}\sigma }{{d\Omega}}({\bm Q})
&=&
\underbrace{\alpha^4
\sum_{\lambda_F} 
\left|{\bm \epsilon}^{\ast}_{{\bm k}_F,\lambda_F} \cdot {\bm \epsilon}_{{\bm k}_{\mathrm{in}},\lambda_{\mathrm{in}}}\right|^2}_
{=: \left(\frac{{\rm d}\sigma }{{\rm d}\Omega}\right)_T}
\times
\left|f^0({\rm Q})\right|^2\;.
\end{eqnarray}
 We defined the elastic Thomson scattering cross section $\left(\frac{{\rm d}\sigma }{{\rm d}\Omega}\right)_T= 0.67 b$ that is a measure of the scattering strength. For light atoms, for which the absorption cross section at threshold is of the order of 1 Mb, the elastic scattering cross section is thus 6 orders of magnitude smaller. \\
\indent Within perturbation theory, that assumes a single scattering event  ({\it kinematic diffraction}), elastic scattering thus gives the square of the Fourier transform of the ground-state electron density. This relation is the underlying basis for structure determination by means of elastic x-ray scattering. Since only the absolute value squared of the Fourier transform is measured, the phase of the elastic scattering factor, that is complex, is lost. This fact is termed {\it phase problem} in crystallography and coherent scattering of noncrystalline samples. In order to determine the electronic structure, the phases have to be reconstructed (see for example \cite{Hauptmann} for Bragg scattering and \cite{Miao} for the method of oversampling of noncrystalline objects).\\
In case of thick samples or high-intensity radiation, multiple interactions with the photon field and multiple scattering events occur, such that the kinematic diffraction theory breaks down. In that case, dynamical diffraction theory \cite{ewald1917,batterman,borrmann} has to be applied, that relies on the lowest order perturbation theory at each particular atomic site of the crystal, but accounts for interference and propagation effects. 
\section{Inelastic X-ray scattering}
In inelastic x-ray scattering processes, the outgoing photon of mode $({\bm k}_{F},\lambda_F)$ has an energy different from the incoming photon $({\bm k}_{\mathrm{in}},\lambda_{\mathrm{in}})$,  leaving the electronic system in a quantum state that is different from the initial state. Starting from the electronic ground state, the electronic/molecular system will be in an excited state following the x-ray interaction, and the outgoing (scattered) electron features an energy loss.  As in the discussion of elastic scattering, we choose the initial state according to eq.\ (\ref{initialstate}), i.e.\ the electronic ground state and the incoming photon field is in a single mode and given by a Fock state. In general the final state of the system therefore reads
\begin{equation}\label{finalinelastsic}
    |F\rangle = 
    |\Psi_F \rangle \hat{a}^\dag_{{\bm k}_{F},\lambda_F}  
    |N_{\mathrm{EM}}-1\rangle\;,
\end{equation}
where we already assumed the lowest order perturbation theory.
The transition to the final state can be mediated in two ways, as in the case of elastic scattering: In first order perturbation theory, the $\hat{A}^2$ can induce the transition by a quasi simultaneous absorption and emission of the photon ("contracted vertex"), similarly to elastic x-ray scattering. Inelastic x-ray scattering by the $\hat{A}^2$ is usually referred to as Thomson Scattering and is the leading contribution far from x-ray absorption edges and resonances. The $\hat{A}\cdot\hat{p}$ term in second order perturbation theory can also lead to inelastic x-ray scattering, often referred to as X-ray Raman Scattering. It is the main contribution of the inelastic scattering cross section close to resonances (resonant inelastic x-ray scattering) or absorption edges. In principle the scattering cross section is a coherent sum of both contributions. In some cases, this can lead to interference effects, that are reflected in specific line shapes of the scattered x-ray spectrum.
According to eq.\ (\ref{gamma}), the general expression for the transition rate from the ground state $|I\rangle$ to the final state $|F\rangle$ is, up to second order contributions in $\hat{A}$ given by 
\begin{eqnarray}
\label{inelastic}
\Gamma_{FI} & = & 
2\pi \delta(E^{N_{\rm el}}_F+\omega_F - E^{N_{\rm el}}_0-\omega_{\rm{in}})\alpha^4\nonumber\\
&\times&\left|
\frac{1}{2} \int {\rm d}^3{\bm x} \;
\langle\Psi_F|\langle N_{\mathrm{EM}}-1|\hat{a}_{{\bm k}_{F},\lambda_{F}}
\hat{\psi}^{\dag}({\bm x})
\hat{A}^2({\bm x})\hat{\psi}({\bm x})|\Psi_0^{N_{\mathrm{el}}}\rangle|N_{\mathrm{EM}}\rangle \right . \nonumber\\
&+& \int {\rm d}^3{\bm x}\int {\rm d}^3{\bm x}'
\sum_{M} \left[\langle\Psi_F|\langle N_{\mathrm{EM}}-1|\hat{a}_{{\bm k}_{F},\lambda_{F}}  \hat{\psi}^{\dag}({\bm x})\hat{A}({\bm x})\cdot\frac{\nabla}{i}\hat{\psi}({\bm x})\right.\nonumber\\
&\times&
\underbrace{
\frac{| \Psi_\mathrm{M}\rangle |N_{\mathrm{EM}}-1\rangle 
\langle N_\mathrm{EM}-1| \langle \Psi_\mathrm{M}| } 
{E^{N_{\rm el}}_0 - E^{N_{\rm el}}_M+\omega_{\rm{in}}
 + {\rm i} \varepsilon}}_{\mbox{regular pathway}}\left.
\hat{\psi}^{\dag}({\bm x'})\hat{A}({\bm x'})\cdot\frac{\nabla}{i}
\hat{\psi}({\bm x'})
|\Psi_0^{N_{\mathrm{el}}}\rangle|N_{\mathrm{EM}}\rangle\right]\nonumber\\
&+& \int {\rm d}^3{\bm x}\int {\rm d}^3{\bm x}'
\sum_{M,m}\left[ \langle\Psi_F |\langle N_{\mathrm{EM}}-1|\hat{a}_{{\bm k}_{F},\lambda_{F}}  \hat{\psi}^{\dag}({\bm x})\hat{A}({\bm x})\cdot\frac{\nabla}{i}\hat{\psi}({\bm x})\right.\nonumber\\
&\times&\underbrace{
\frac{ |\Psi_\mathrm{M}\rangle \hat{a}^\dag_{{\bm k}_{m},\lambda_{m}}| N_{\mathrm{EM}}\rangle 
\langle N_\mathrm{EM}|\hat{a}_{{\bm k}_{m},\lambda_{m}}| \langle \Psi_\mathrm{M}| }
{E^{N_{\rm el}}_0 - E^{N_{\rm el}}_M-\omega_{m}
+{\rm i} \varepsilon}}_{\mbox{irregular pathway}}
\left.\left.
 \hat{\psi}^{\dag}({\bm x'})\hat{A}({\bm x'})\cdot\frac{\nabla}{i}
\hat{\psi}({\bm x'})
|\Psi_0^{N_{\mathrm{el}}}\rangle|N_{\mathrm{EM}\rangle}\right]
\right|^2
\end{eqnarray}
In applying eq.\ (\ref{gamma}) care has to be taken in the second order term: The sum over intermediate states appearing in eq.\ (\ref{gamma}) implies the sum over both, electronic and field degrees of freedom. Keeping in mind, that in the second order perturbation theory, the initial and intermediate states are connected by a linear interaction in the vector potential $\hat{A}$, only intermediate states of the form $|N_\mathrm{EM}-1\rangle$ and $\hat{a}^\dag_{{\bm k}_{m},\lambda_{m}}|N_\mathrm{EM}\rangle$ that differ from the initial photon state vector by either the application of an annihilation in the incoming mode or a creation operator in any other mode contribute to the sum over the intermediate states. Therefore, the $\hat{A}\cdot p$ contribution has in principle two pathways: the regular pathway and the irregular pathway, where photo emission of the scattered photon happens before the annihilation of a photon in the incoming mode. This quasi "time-reversed" process is typically negligible at resonances \cite{rixs}. The differential scattering cross section for inelastic x-ray scattering that follows from eq.\ (\ref{inelastic}) is typically referred to as the Kramers-Heisenberg equation \cite{kramers}.\\
\indent Eq. \ref{inelastic} also describes the non relativistic limit of Compton scattering, which is mediated by the $A^2$-term supposing an electron that is initially at rest. A derivation of the non-relativistic Compton scattering cross section is given in reference \cite{santra} and shall not be repeated here. In contrast, we focus on the inelastic scattering contributions that have not been discussed in \cite{santra}.\\
We now suppose that the different pathways of inelastic scattering of eq.\ (\ref{inelastic}) do not interfere, and look at the individual contributions independently.
\subsection{X-ray Thomson scattering}
\indent  We first derive the double-differential inelastic scattering cross section induced by $\hat{A}^2$. Along similar lines as the derivation of the elastic scattering transition rate for the $\hat{A}^2$ contribution, we get the transition rate from initial state $|I\rangle$ of eq.\ (\ref{initialstate}) to final state $|F\rangle$ of eq.\ (\ref{finalinelastsic}):
\begin{eqnarray}
\label{inelasticThomson}
\Gamma_{FI}  &=&  
2\pi \delta(E_F - E_I)\frac{\alpha^4}{4}
 \left|\int {\rm d}^3{\bm x} \; 
 \langle\Psi_F|\langle N_{\mathrm{EM}}-1|\hat{a}_{{\bm k}_{F},\lambda_{F}}
\hat{\psi}^{\dag}({\bm x}) \hat{A}^2({\bm x})\hat{\psi}({\bm x})
|\Psi_0^{N_{\mathrm{el}}}\rangle|N_{\mathrm{EM}}\rangle \right |^2\nonumber\\
&=&
\frac{(2\pi)^3}{V}\frac{N_{EM}}{V}\frac{1}{\omega_F\omega_{in}} 
\delta(E_F^{N_{\rm{el}}}- E_0^{N_{\rm{el}}}   +\omega_F-\omega_{\rm in})
\left|{\bm \epsilon}^{\ast}_{{\bm k}_F,\lambda_F} \cdot {\bm \epsilon}_{{\bm k}_{\mathrm{in}},\lambda_{\mathrm{in}}} \right|^2\nonumber\\
&&\;\;\;\;\;\;\;\;\;\;\;\;\;\;\;\;\;\;\;\;\;\;\;\;\;\;\;\;\;\;\;\;\;\;\;\;
\times\left|\int {\rm d}^3{\bm x}\; 
\langle\Psi_\mathrm{F}^{N_{\rm{el}}}|\hat{\psi}^{\dag}({\bm x}){\rm e}^{{\rm i}({\bm k}_{\mathrm{in}}-{\bm k}_F)\cdot{\bm x}}
\hat{\psi}({\bm x})|\Psi_0^{N_{\mathrm{el}}}\rangle\right|^2\;.
\end{eqnarray}
The difference here to the case of elastic scattering  is the that the final state of matter is generally not equal to the initial state. Similarly to elastic scattering the matrix element involves the momentum transfer $\bm Q=\bm k_{in}- \bm k_F$ in the exponential, the momentum transfer, however, goes along with an energy transfer $\Delta \omega := \omega_{\rm in}-\omega_F$ that is generally unequal to zero.\\
\indent In a typical inelastic x-ray scattering experiment, the scattered photons are detected along a specific outgoing $\bm k$--vector $\hat{\bm k}_F=\bm{\Omega}$ within a solid angle $\bm \Delta \bm \Omega$. Whereas in  typical coherent x-ray scattering experiment aiming for structure determination the photon energy remains unobserved, in inelastic x-ray scattering experiments, the photon energy is resolved. To derive a differential cross section, we have to integrate over the experimentally unobserved degrees of freedom and divide by the incoming photon flux. The polarisation of the scattered light typically stays unobserved. To obtain a differential cross section, the transition rate is therefore summed over the polarisation $\lambda_F$, and summed over a finite width ${\bm \Delta}{\bm \Omega}$ of the outgoing $\bm k$-vector of photon momentum $|k_F|$. The sum over photon momenta translates into an integral over the normalised outgoing $\bm k$-vector $\bm \Omega$ and the photon energy $\omega_F$ of the emitted photon:
\begin{equation}
\sum_{{\bm \Delta \bm \Omega}, |k_F|-\delta<|k_F|<|k_F|+\delta} =\alpha^3 \frac{V}{(2\pi)^3}
\;{\rm d}{\bm \Omega} \;{\rm d}\omega_F\;\omega_F^2
\end{equation}
Since the state of the electronic/molecular system remains unresolved in the experiment, we also have to sum over all possible final states $|\Psi_F^{N_{\rm el}}\rangle$.
Dividing the transition rate by the incoming photon flux defined in eq.\ (\ref{flux}) then leads to the double differential inelastic Thomson
scattering cross section
\begin{eqnarray}
\label{ddcs1}
\frac{{\rm d}^2\sigma}{{\rm d}{\bm \Omega}{\rm d}\omega_F}({\bm Q},\omega_{\rm in}) 
&=& 
\alpha^4\frac{\omega_F}{\omega_{\mathrm{in}}}
\sum_{\lambda_F} 
\left|{\bm \epsilon}^{\ast}_{{\bm k}_F,\lambda_F} \cdot {\bm \epsilon}_{{\bm k}_{\mathrm{in}},\lambda_{\mathrm{in}}}\right|^2\nonumber\\
&\times&
\underbrace{\sum_{F}
\delta(E_F^{N_{\rm{el}}}- E_0^{N_{\rm{el}}}-\omega_{\rm i}+\omega_F)
\left|\int {\rm d}^3{\bm x} \langle\Psi_\mathrm{F}|\hat{\psi}^{\dag}({\bm x}){\rm e}^{{\rm i}{\bm Q}\cdot {\bm x}}
\hat{\psi}({\bm x})|\Psi_0^{N_{\mathrm{el}}}\rangle\right|^2}_{=: A}.
\end{eqnarray}
In order to give more physical insight into inelastic x-ray scattering, we rewrite eq.\ {\ref{ddcs1}} by expressing the delta function for energy conservation as a time integral:
\begin{equation}
\label{last}
    \delta(E_F^{N_{\rm{el}}}- E_0^{N_{\rm{el}}}+\omega_F-\omega_{\rm in})=
    \frac{1}{2\pi}\int {\rm d}t \;\;
    {\rm e} ^{{\rm i} (E_F^{N_{\rm{el}}}- E_0^{N_{\rm{el}}}+\omega_F-\omega_{\rm in}) t}
\end{equation}
With $\Delta\omega:=\omega_{\rm in}-\omega_F$, the second line of eq.\ (\ref{ddcs1}) becomes
\begin{eqnarray}
A&=&\frac{1}{2\pi}\int {\rm d}t \;{\rm e}^{- {\rm i} \Delta\omega t} \sum_F  
\int {\rm d}^3{\bm x}' \int {\rm d}^3{\bm x}\;
\langle\Psi_0^{N_{\mathrm{el}}}|\hat{\psi}^\dag({\bm x'})
{\rm e}^{-{\rm i}{\bm Q\cdot \bm x'}}
\hat{\psi}({\bm x'})|\Psi_\mathrm{F}^{N_{\rm{el}}}\rangle\nonumber\\
&&\times\langle\Psi_\mathrm{F}^{N_{\rm{el}}}| {\rm e}^{{\rm i} E_F^{N_{\rm{el}}} t}
\hat{\psi}^{\dag}({\bm x}){\rm e}^{{\rm i}{\bm Q}\cdot{\bm x}}
\hat{\psi}({\bm x}) 
{\rm e}^{-{\rm i} E_0^{N_{\rm{el}}} t}|\Psi_0^{N_{\mathrm{el}}}\rangle
\end{eqnarray}
${\rm e}^{-{\rm i} E_0^{N_{\rm{el}}} t}|\Psi_0^{N_{\mathrm{el}}}\rangle$ can be rewritten as
${\rm e}^{-{\rm i} \hat{H}_{\rm{el}} t}|\Psi_0^{N_{\mathrm{el}}}\rangle$, and likewise $\langle\Psi_\mathrm{F}| {\rm e} ^{{\rm i} E_F^{N_{\rm{el}}} t}= \langle\Psi_\mathrm{F}|{\rm e}^{{\rm i} \hat{H}_{\rm{el}} t}$. Then, realising that the electronic density operator in the interaction picture is given by
\begin{equation}
\hat{n}({\bm x}, t) = {\rm e}^{{\rm i} \hat{H}_{\rm{el}} t}\hat{\psi}^{\dag}({\bm x})\hat{\psi}({\bm x}){\rm e}^{-{\rm i} \hat{H}_{\rm{el}} t}
\end{equation}
and making use of the resolution of identity that eliminates the sum over final states $\Psi_F$, we finally find
\begin{eqnarray}
\label{ddcs2}
\frac{{\rm d}^2\sigma}{{\rm d}{\bm \Omega}{\rm d}\omega_F}({\bm Q},\omega_{\rm in})&=& 
\alpha^4\frac{\omega_F}{\omega_{\mathrm{in}}}
\left|{\bm \epsilon}^{\ast}_{{\bm k}_F,\lambda_F} \!\!\!\!\cdot {\bm \epsilon}_{{\bm k}_{\mathrm{in}},\lambda_{\mathrm{in}}}\right|^2\nonumber\\
&&\times
\int \frac{{\rm d}t}{2\pi}
{\rm e} ^{-{\rm i}\Delta\omega t}
\int\!\!\!\!\!\int {\rm d}^3{ \bm x}' {\rm d}^3{\bm x}\; 
\langle \Psi_0^{N_{\mathrm{el}}}|
\hat{n}({\bm x'},0)
{\rm e}^{{\rm i}{\bm Q}\cdot ({ \bm x}-{\bm x}')}
\hat{n}({\bm x},t)|\Psi_0^{N_{\mathrm{el}}}\rangle
\end{eqnarray}
The inelastic scattering cross section therefore measures electron density-density correlations of a specific momentum transfer vector $\bm Q$ in time.
\\
\subsection{Dynamic structure factor of inelastic x-ray scattering}
\indent We now will derive an adiabatic approximation of the inelastic double differential scattering cross section of eq.\ (\ref{ddcs2}) within the independent atom model. This is an approximation that is widely used for the interpretation of inelastic x-ray scattering of disordered materials and relates the scattering cross section to the pair-distribution function.
For simplicity, we assume a single atomic species of the many-body system under consideration. We treat the atomic nuclei as an ensemble of classical particles that move on trajectories ${\bm R}_n(t)$. The electrons pertaining to a specific nucleus follow that motion adiabatically and we assume that the atoms are in their ground states with a ground-state density $\rho_0^{\rm el}({\bm x - \rm R}_n(t))$. This approximation is valid for an energy transfer $\Delta\omega$ in Eq.\ (\ref{ddcs2}) that is small compared to typical electronic excitation energies. The Fourier integral $\int \frac{{\rm d}t}{2\pi} {\rm e} ^{-{\rm i}\Delta\omega t}$ in Eq.\ (\ref{ddcs2}) in that case does not resolve electron dynamics and fluctuations in the electronic density-density correlation function. The energy range of excitations that qualify for this adiabatic approximation are thus in the range of a typical phonon excitation. By restricting the energy transfer to be small compared to the electronic excitation energies, the electronic system remains unchanged during the scattering process. The difference between initial and final state is thus purely associated with atomic density fluctuations, that we treat classically.
Within this adiabatic approximation the electron density operator can be rewritten as a sum over single-atom density operators centred around ${\bm R}_n$:
\begin{equation}
    \hat{n}({\bm r};t)= \sum_{n=1}^N \hat{n}[{\bm x- R}_n (t)]\;,
\end{equation}
where the sum runs over all $N$ atoms of the system. Under this approximation Eq.\ (\ref{ddcs2}) becomes
\begin{eqnarray}
\label{ddcs3}
\frac{{\rm d}^2\sigma({\bm Q})}{{\rm d}{\bm \Omega}{\rm d}\omega_F} &=& \alpha^4\frac{\omega_F}{\omega_{\mathrm{in}}}
\sum_{\lambda_F} 
\left|{\bm \epsilon}^{\ast}_{{\bm k}_F,\lambda_F} \cdot {\bm \epsilon}_{{\bm k}_{\mathrm{in}},\lambda_{\mathrm{in}}}\right|^2
\frac{1}{2\pi}\int {\rm d}t\;
{\rm e} ^{- {\rm i} \Delta\omega t}
\nonumber\\
&\times&
\sum_{n,n'}
\int \!{\rm d}^3{\bm x}' \!\!\int\! {\rm d}^3{\bm x} \;
\langle \Psi_0^{N_{\mathrm{el}}}|
\hat{n}[{\bm x'-\bm R}_{n'}(0)]
{\rm e}^{{\rm i}{\bm Q}\cdot ({ \bm x-\bm x'})}
\hat{n}[{\bm x-\bm R}_n(t)]|\Psi_0^{N_{\mathrm{el}}}\rangle
\end{eqnarray}
Note, that the electronic state vector in Eq.\ ($\ref{ddcs2}$) corresponds to the correlated total molecular many-body wave function, here, since we assumed the independent atom model $|\Psi_0^{N_{\mathrm{el}}}\rangle$ stands for the (many-body) wave function of a single atom. Moreover we assume, that electronic state remains unchanged by the scattering event.
With the variable substitutions $\bm x \rightarrow \bm x - \bm R_n$ and  $\bm x' \rightarrow \bm x' - \bm R_{n'}$ we get
\begin{eqnarray}
\label{ddcs4}
\frac{{\rm d}^2\sigma({\bm Q})}{{\rm d}\Omega{\rm d}\omega_F} &=& \alpha^4\frac{\omega_F}{\omega_{\mathrm{in}}}
\sum_{\lambda_F} 
\left|{\bm \epsilon}^{\ast}_{{\bm k}_F,\lambda_F} \cdot {\bm \epsilon}_{{\bm k}_{\mathrm{in}},\lambda_{\mathrm{in}}}\right|^2
\nonumber\\
&&\times
\underbrace{\int {\rm d}^3{\bm x}' \!\!\int {\rm d}^3{\bm x} \;
\rho^{el}_0 ({\bm x'})
{\rm e}^{{\rm i}{\bm Q}\cdot (\bm x- \bm x')}
\rho^{el}_0 ({\bm x})}_
{=|f_0({\bm Q})|^2}
\times
\underbrace{
\frac{1}{2\pi}\int {\rm d}t\;
{\rm e} ^{- {\rm i} \Delta\omega t}
\sum_{n,n'}
{\rm e}^{{\rm i}{\bm Q}\cdot ( \bm R_n(t)-\bm R_{n'}(0))}}
_{S({\bm Q},\Delta\omega)}
\end{eqnarray}
In the adiabatic, independent atom approximation the double differential cross section therefore factorises
into the form factor squared of a single atom $f_0(\bm Q)$  and a  point-particle density-density correlation in space and time -- the so called dynamic structure factor $S(\bm{Q},\Delta\omega)$ of momentum transfer $\bm{Q}$ and energy transfer $\Delta\omega$ :
\begin{equation}
  S(\bm Q,\Delta\omega)=
  \frac{1}{2\pi}\int {\rm d}t\; {\rm e} ^{-{\rm i}\Delta\omega t}
\sum_{n,n'} {\rm e}^{{\rm i}{\bm Q}\cdot (\bm R_n(t)-\bm R_{n'}(0))}\;.
\end{equation}
The dynamical structure factor is directly linked to the time-dependent pair-distribution function of  disordered materials  \cite{vanhove} by
\begin{eqnarray}
G(\bm R,t)&=&\frac{1}{(2\pi)^3} \frac{1}{N}
\int {\rm d}^3\bm Q \int {\rm d}\Delta\omega \;
{\rm e}^{{\rm i}(\Delta\omega t -\bm Q\cdot \bm R)} S(\bm Q,\Delta\omega)\;,
\end{eqnarray}
where $N$ is equal to the sum of atoms in the system.
The inelastic x-ray scattering cross section therefore can be rewritten as
\begin{eqnarray}
\label{ddcs}
\frac{{\rm d}^2\sigma({\bm Q})}{{\rm d}\bm \Omega{\rm d}\omega_F}=
\alpha^4\frac{\omega_F}{\omega_{\mathrm{in}}}
\sum_{\lambda_F} 
\left|{\bm \epsilon}^{\ast}_{{\bm k}_F,\lambda_F} \cdot {\bm \epsilon}_{{\bm k}_{\mathrm{in}},\lambda_{\mathrm{in}}}\right|^2
 |f_0(\bm {Q})|^2\times S({\bm Q},\Delta \omega)\;.
\end{eqnarray}
Whereas elastic x-ray scattering measures the static density distribution, inelastic scattering gives access to particle correlations in space and time, and the time-dependent pair distribution function. Hence, inelastic x-ray scattering, and similarly inelastic neutron scattering, are the core experimental techniques to study liquids, glasses \cite{sette} and liquid metals \cite{Scopigno}.\\
\subsection{Resonant Inelastic X-ray Scattering (RIXS)}
Now we consider the inelastic scattering contribution due to the $\hat{A}\cdot\hat{p}$ contribution in second order of eq.\ (\ref{inelastic}) to the final state specified in eq.\ (\ref{finalinelastsic}). The transition rate up to second order in perturbation theory reads:
\begin{eqnarray}
\Gamma_{FI} & = & 
2\pi \delta(E^{N_{\rm el}}_F+\omega_F - E^{N_{\rm el}}_0-\omega_{\rm{in}})\alpha^4\nonumber\\
&\times&\left|
 \int {\rm d}^3{\bm x}\int {\rm d}^3{\bm x}'
\sum_{M} 
\langle\Psi_F^{N_{\mathrm{el}}}|\langle N_{\mathrm{EM}}-1|\hat{a}_{{\bm k}_{F},\lambda_{F}}  \hat{\psi}^{\dag}({\bm x})\hat{A}({\bm x})\cdot\frac{\bm \nabla}{i}\hat{\psi}({\bm x})\right.\nonumber\\
&\times&
\frac{| \Psi_M^{N_{\mathrm{el}}}\rangle |N_{\mathrm{EM}}-1\rangle 
\langle N_\mathrm{EM}-1| \langle \Psi_M^{N_{\mathrm{el}}}| } 
{E^{N_{\rm el}}_0 - E^{N_{\rm el}}_M+\omega_{\rm{in}}
 + {\rm i} \varepsilon}\left.
\hat{\psi}^{\dag}({\bm x'})\hat{A}({\bm x'})\cdot\frac{\bm \nabla'}{i}
\hat{\psi}({\bm x'})
|\Psi_0^{N_{\mathrm{el}}}\rangle|N_{\mathrm{EM}}\rangle
\right|^2
\end{eqnarray}
Collecting all the nonzero terms in the expansion of the vector potential in terms of annihilation and creation operators leads to
\begin{eqnarray}
\Gamma_{FI} & = & 
2\pi \delta(E^{N_{\rm el}}_F+\omega_F - E^{N_{\rm el}}_0-\omega_{\rm{in}}) 
\frac{(2\pi)^2N_{\mathrm{EM}}}{V^2\omega_F\omega_{\rm{in}}}
\nonumber\\
&\times&\left|
 \int {\rm d}^3{\bm x}\int {\rm d}^3{\bm x}'
\sum_{M} 
\langle\Psi_F^{N_{\mathrm{el}}}|  \hat{\psi}^{\dag}({\bm x})
{\rm e}^{-{\rm i}{\bm k_F}\cdot{\bm x}}
{\bm \epsilon}^{\ast}_{{\bm k_F},\lambda_F}\cdot\frac{\bm \nabla}{\rm i}\hat{\psi}({\bm x})\right.\nonumber\\
&\times&\left.
\frac{| \Psi_M^{N_{\mathrm{el}}}\rangle  
\langle \Psi_M^{N_{\mathrm{el}}}| } 
{E^{N_{\rm el}}_0 - E^{N_{\rm el}}_M+\omega_{\rm{in}}
 + {\rm i} \varepsilon}
\hat{\psi}^{\dag}({\bm x'})
{\rm e}^{{\rm i}{\bm k_{\rm{in}}}\cdot{\bm x'}}
{\bm \epsilon}_{{\bm k_{\bm{in}}},\lambda_{\rm{in}}} \cdot
\frac{\bm \nabla'}{\rm i}
\hat{\psi}({\bm x'})
|\Psi_0^{N_{\mathrm{el}}}\rangle
\right|^2
\end{eqnarray}
To interpret RIXS in terms of single particle-hole transitions , we adopt the expansion of the field operators in terms of spin orbitals of an appropriate mean-field Hamiltonian according to Eq.\ \ref{psiexp}:
\begin{eqnarray}
\Gamma_{FI} & = & 
2\pi \delta(E^{N_{\rm el}}_F+\omega_F - E^{N_{\rm el}}_0-\omega_{\rm{in}}) 
\frac{(2\pi)^2N_{\mathrm{EM}}}{V^2\omega_F\omega_{\rm{in}}}
\nonumber\\
 &\times&\left| \sum_{i,a,b,p,M}\int {\rm d}^3{\bm x} \varphi_p^{\dag}({\bm x})
 {\rm e}^{-{\rm i}{\bm k_F}\cdot{\bm x}}{\bm \epsilon}^{\ast}_{{\bm k_F},\lambda_F}\cdot\frac{\bm \nabla}{\rm i}
\varphi_b({\bm x})
 \int {\rm d}^3{\bm x}' \varphi_a^{\dag}({\bm x'}) 
{\rm e}^{{\rm i}{\bm k_{\rm{in}}}\cdot{\bm x'}}
{\bm \epsilon}_{{\bm k_{\bm{in}}},\lambda_{\rm{in}}}\frac{\bm \nabla'}{\rm i}
\varphi_i({\bm x'})\right.\nonumber\\
&&\;\;\;\;\;\;\;\;\;\;\;\;\;\;\;\;\;\;\;\;\;\;\;\;
\times\left.
\frac{\langle\Psi_F^{N_{\mathrm{el}}}|\hat{c}^{\dag}_p\hat{c}_b | \Psi_M^{N_{\mathrm{el}}}\rangle
\langle \Psi_M^{N_{\mathrm{el}}}| \hat{c}^{\dag}_a\hat{c}_i |\Psi_0^{N_{\mathrm{el}}}\rangle}
 {E^{N_{\rm el}}_0 - E^{N_{\rm el}}_M+\omega_{\rm{in}}
 + {\rm i} \varepsilon}
\right|^2
\end{eqnarray}
In this expression, the initial, intermediate and final electronic states are in principle fully-correlated wave functions. To facilitate the interpretation, we now adopt the mean-field model also for the initial state. As in the case for photo absorption, the ground-state wave function is replaced by the ground-state Slater determinant of the system, so that absorption of a photon promotes an initially occupied orbital $i$ to an unoccupied orbital $a$: $ \hat{c}^{\dag}_a\hat{c}_i |\Psi_0^{N_{\mathrm{el}}}\rangle\approx  \hat{c}^{\dag}_a\hat{c}_i |\Phi_0^{N_{\mathrm{el}}}\rangle  = |\Phi_i^a\rangle$. We adopt the somewhat sloppy notation of the single-particle transition matrix elements in terms of bra-ket notation. Approximating the ground-state by the Slater determinant yields
\begin{eqnarray}\label{rixs2}
\Gamma_{FI} & = & 
2\pi \delta(E^{N_{\rm el}}_F+\omega_F - E^{N_{\rm el}}_0-\omega_{\rm{in}}) 
\frac{(2\pi)^2N_{\mathrm{EM}}}{V^2\omega_F\omega_{\rm{in}}}
\nonumber\\
 &\times&\left| \sum_{i,a,b,p,M}
\langle\varphi_p|
 {\rm e}^{-{\rm i}{\bm k_F}\cdot{\bm x}}{\bm \epsilon}^{\ast}_{{\bm k_F},\lambda_F}\cdot\frac{\nabla}{\rm i}
|\varphi_b\rangle 
 \langle \varphi_a|
{\rm e}^{{\rm i}{\bm k_{\rm{in}}}\cdot{\bm x'}}
{\bm \epsilon}_{{\bm k_{\bm{in}}},\lambda_{\rm{in}}}\frac{\nabla'}{\rm i}
|\varphi_i\rangle\right.\nonumber\\
&&\;\;\;\;\;\;\;\;\;\;\;\;\;\;\;\;\;\;\;\;\;\;\;\;
\times\left.
\frac{\langle\Psi_F^{N_{\mathrm{el}}}|\hat{c}^{\dag}_p\hat{c}_b | \Psi_M^{N_{\mathrm{el}}}\rangle
\langle \Psi_M^{N_{\mathrm{el}}}|\Phi_i^a\rangle}
 {E^{N_{\rm el}}_0 - E^{N_{\rm el}}_M+\omega_{\rm{in}}
 + {\rm i} \varepsilon}
\right|^2\;.
\end{eqnarray}
\begin{figure}
    \centering
    \includegraphics[width=8cm]{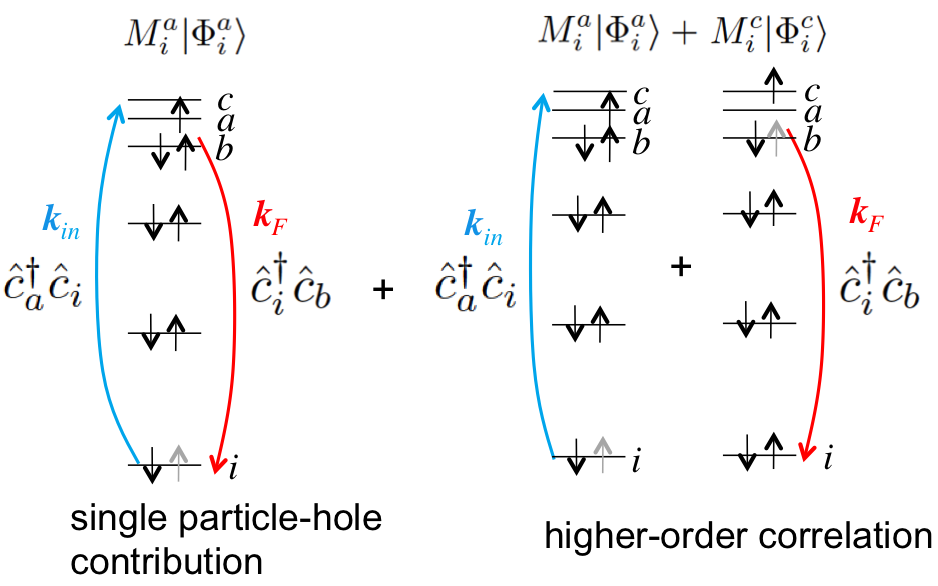}
    \caption{Schematic level scheme of the RIXS process.}
    \label{fig:3}
\end{figure}

At this point, the sum over intermediate states in principle runs over all possible correlated states. We now expand the indermediate states in terms of a configuration interaction in the single-particle basis:
\begin{equation}\label{corrinter}
    |\Psi_M^{N_{\mathrm{el}}}\rangle= M_0 |\Phi_0\rangle+ \sum_{i,a}M_i^a|\Phi_i^a\rangle +\sum_{j,k,c,d}M_{j,k}^{c,d}|\Phi_{j,k}^{c,d}\rangle +....
\end{equation}
where we introduced the two-particle two-hole functions 
$|\Phi_{j,k}^{c,d}\rangle:= \hat{c}^\dag_d\hat{c}^\dag_c \hat{c}_k\hat{c}_j|\Phi_{0}\rangle$ 
and the corresponding expansion coefficients. The overlap matrix element gives $\langle\Psi_M^{N_{\mathrm{el}}}|\Phi_i^a\rangle=(M_i^a)^*$. The first interaction with the photon field thus probes single particle-hole excitations of the ground-state. The sum over intermediate states, however, extends over all intermediate states that are reasonably close in energy to the incoming photon. Energy is strictly speaking not conserved in this excitation step, only the overall energy has to be conserved, which is enforced by the delta function. In practice, the energy-denominator suppresses contributions from intermediate states that are "off-shell". To further understand the contributing interaction terms, let us suppose that there is only a single intermediate $M$ state that is reachable by the incoming photon energy, but that features strong configuration interaction, as expressed in Eq.\ (\ref{corrinter}). The second interaction with the vector potential then results in an emission of a photon of energy $\omega_F$ by annihilation of an electron in orbital $b$ and occupying a deeper-lying orbital $p$. The process is depicted schematically in Figure \ref{fig:3}. For every practical purpose, starting from the ground-state and for photon energies above inner-shell ionisation edges $p=i$. This means, the initially created deep core hole is refilled in the RIXS process. The orbital $b$ has to be occupied in the intermediate state. The strongest contribution will be given by dexcitation of a state $b$ that is initially occupied in the ground state. But since the intermediate state can show configuration interaction, other, non-direct
terms can contribute. This is depicted in Figure \ref{fig:3}, where a contribution to the RIXS signal is shown, that shows higher-order correlation effects. In the depicted case, for example, the intermediate state shows configuration mixing in the single-particle hole expansion and additional, correlation driven terms, contribute to the transition rate. This discussion should highlight that  RIXS probes electron correlation, despite the fact that the system is probed by a photon in / photon out process, that in an intuitive view based on the single-particle basis function expansion, can be described in terms of single-particle hole excitations and de-excitations. Along this line of argumentation, it shall be noted that RIXS, in addition to probing elementary electronic excitations \cite{kotani}, also probes collective excitation modes, such as phonons, charge-transfer modes, etc.  in solids. Since the involved matrix elements also include the $\bm k$-vector of the in- and outgoing photon field, RIXS also probes the momentum of the elementary excitations and in addition to Thomson scattering is therefore a viable and powerful tool to study elementary excitations in solids. For more information see for example reference \cite{rixs,ament}. High-resolution RIXS, measuring small energy transfer in the range of meV also opens the possibility to study elementary excitations in superconductivity and is a vivid research field \cite{baron1,baron2}.\\
\indent As discussed in the last paragraph, RIXS is a coherent scattering process, that cannot be partitioned in cross sections of a photoabsorption followed by a photoemission process. A particular feature of a resonance scattering process is the {\it anomalous} linear dispersion relation of the outgoing photon energy. To highlight this fact, we now restrict our underlying electronic model system to an uncorrelated electronic system, that can purely be described by single-particle hole configurations, in the initial, intermediate and final state. We now suppose that only a single intermediate state can be reached with an incoming photon energy $\omega_{\mathrm{in}}$ state in Eq.\ (\ref{rixs2}):  $|\Psi_M^{N_{\mathrm{el}}}\rangle= |\Phi_i^a\rangle$. The final state shall be approximated by $|\Psi_F^{N_{\mathrm{el}}}\rangle=|\Phi_b^a\rangle$. The transition rate then takes the simplified form
\begin{eqnarray}\label{rixs3}
\Gamma_{ab} & = & 
2\pi \delta(\epsilon_a-\epsilon_b +\omega_F -\omega_{\rm{in}}) 
\frac{(2\pi)^2N_{\mathrm{EM}}}{V^2\omega_F\omega_{\rm{in}}}
\nonumber\\
 &\times&\left|
\frac{\langle\varphi_i|
 {\rm e}^{-{\rm i}{\bm k_F}\cdot{\bm x}}{\bm \epsilon}^{\ast}_{{\bm k_F},\lambda_F}\cdot\frac{\bm \nabla}{\rm i}
|\varphi_b\rangle 
 \langle \varphi_a|
{\rm e}^{{\rm i}{\bm k_{\rm{in}}}\cdot{\bm x'}}
{\bm \epsilon}_{{\bm k_{\bm{in}}},\lambda_{\rm{in}}}\frac{\bm \nabla'}{\rm i}
|\varphi_i\rangle}
{\epsilon_i-\epsilon_a +\omega_{\rm{in}}
 + {\rm i} \varepsilon}
\right|^2\;.
\end{eqnarray}
The partial differential cross section related to the final state $\Phi_b^a$ is then given by
\begin{eqnarray}
\frac{{\rm d}^2 \sigma_{ab}}{{\rm d}\Omega{\rm d}\omega_{F}} & = & 
\delta(\epsilon_a-\epsilon_b +\omega_F -\omega_{\rm{in}}) 
\alpha^4 \frac{\omega_F}{\omega_{\rm{in}}}
\nonumber\\
 &\times&
 \left|
\frac{\langle\varphi_i|
 {\rm e}^{-{\rm i}{\bm k_F}\cdot{\bm x}}{\bm \epsilon}^{\ast}_{{\bm k_F},\lambda_F}\cdot\frac{\bm \nabla}{\rm i}
|\varphi_b\rangle 
 \langle \varphi_a|
{\rm e}^{{\rm i}{\bm k_{\rm{in}}}\cdot{\bm x'}}
{\bm \epsilon}_{{\bm k_{\bm{in}}},\lambda_{\rm{in}}}\frac{\bm \nabla'}{\rm i}
|\varphi_i\rangle}
{\epsilon_i-\epsilon_a +\omega_{\rm{in}}
 + {\rm i} \varepsilon}
\right|^2\;.
\end{eqnarray}
The small quantity $\varepsilon$ in the denominator, that resulted out of the adiabatic switching of the interaction, can be phenomenologically assigned with the lifetime width of the core-excited intermediate state of the RIXS process. This can be formally derived within the so-called sudden approximation, which is out of the scope of this work. We shall consider $\varepsilon$ as the combined Auger- and radiative width of the intermediate state.
Tuning $\omega_{\mathrm{in}}$ through the resonance $\epsilon_a-\epsilon_i$ we observe two important features. Due to the energy denominator, the cross section would fall off rapidly for a detuning $\delta= \epsilon_a-\epsilon_i-\omega_{\mathrm{in}}$ that is large compared to the width $\varepsilon$ of the resonance. Since the overall energy conservation is between initial and final states, the outgoing photon energy $\omega_F$ depends linearly on the detuning $\delta$ and thus the incoming photon energy $\omega_{\rm in}$, when $\omega_{\rm in}$ is tuned through the resonance. This behaviour is called {\it anomalous} linear energy dispersion of resonance scattering  and a characteristic feature of RIXS. This holds in the limit of narrow-band excitation (bandwidth of source is small as compared the natural resonance width), trivially fulfilled within our approach of a single mode in the incoming photon field. The outgoing radiation in this limit picks up the line width of the incoming radiation, thus emission narrower than the lifetime width can be observed. This behaviour was first demonstrated in 1976 \cite{eisenberger}.
\section{Conclusions}
Within this short introduction to x-ray matter interaction, I gave a concise summary of x-ray matter interaction based on perturbation theory and derived cross-sections for the most important processes, such as photo absorption, elastic scattering and inelastic x-ray scattering. The article is not to be understood as a review article of the field, neither was I trying to give a historic overview of the development of the field. Nevertheless, I hope that the student can build up on this basis and can link the measurable quantities in typical x-ray based experiments to the underlying elementary quantities. I hope that the article gives enough stimulus and ideas of further reading and study, and that is serves as a good starting point, to connect with the other, more specialised lectures of this summer school.\\
\indent I thank Robin Santra for providing the source file of his tutorial \cite{santra}, which tremendously facilitated the writing of these lecture notes. I also would like to thank Andrei Benediktovitch and Dietrich Krebs for the careful reading of the manuscript and  for checking the equations for typos, missing factors or other inconsistencies.

\end{document}